\documentclass[10pt,conference]{IEEEtran}
\IEEEoverridecommandlockouts
\usepackage{amsmath,amssymb,amsfonts}
\usepackage{graphicx}
\usepackage{url}
\usepackage{stfloats}
\usepackage{xcolor}
\usepackage{paralist}
\usepackage{xspace}
\usepackage{hyperref}
\usepackage[inkscapelatex=false]{svg}

\usepackage{tikz}
\usepackage{fontawesome}
\hypersetup{
colorlinks = true, 
hidelinks,
linkcolor=black, 
citecolor=black, 
urlcolor=black 
}

\def\BibTeX{{\rm B\kern-.05em{\sc i\kern-.025em b}\kern-.08em
T\kern-.1667em\lower.7ex\hbox{E}\kern-.125emX}}

\newcommand{\qmutbench}{\texttt{QMutBench}\xspace}
\newcommand{\survivalrate}{\texttt{survival rate}\xspace}
\newcommand{\qasm}{\texttt{QASM}\xspace}

\begin{document}

\title{QMutBench: A Dataset of Quantum Circuit Mutants
\thanks{E. Mendiluze Usandizaga is supported by Simula's internal strategic project on quantum software engineering. P. Arcaini is supported by the ASPIRE grant No. JPMJAP2301, JST. S. Ali is supported by the Qu-Test project (Project \#299827) funded by the Research Council of Norway and Oslo Metropolitan University's Quantum Hub.}
}

\author{\IEEEauthorblockN{Eñaut Mendiluze Usandizaga}
\IEEEauthorblockA{\textit{Simula Research Laboratory and} \\
\textit{Oslo Metropolitan University}\\
Oslo, Norway \\
enaut@simula.no}
\\
\IEEEauthorblockN{Paolo Arcaini}
\IEEEauthorblockA{\textit{National Institute of Informatics}\\
Tokyo, Japan \\
arcaini@nii.ac.jp}
\and
\IEEEauthorblockN{Thomas Laurent}
\IEEEauthorblockA{\textit{Trinity College Dublin, School of Computer Science and Statistics} \\
Dublin, Ireland \\
tlaurent@tcd.ie}
\\
\IEEEauthorblockN{Shaukat Ali}
\IEEEauthorblockA{\textit{Simula Research Laboratory and} \\
\textit{Oslo Metropolitan University}\\
Oslo, Norway \\
shaukat@simula.no}
}

\maketitle

\begin{abstract}
Quantum software testing has attracted interest in recent years, prompting the development of various techniques to automate the testing of quantum software. These techniques generate test cases that must be assessed for their effectiveness in detecting faults. Such an assessment requires benchmarks of faulty programs. However, there is a lack of benchmarks containing faults. In this data showcase, we propose \qmutbench, a dataset that contains over 700,000 quantum circuit mutants representing different faults. The dataset is accessible via an online interface with selection criteria, such as the original quantum circuit(s) from which mutants are generated, the desired survival rate of the selected mutants, and other mutation characteristics (e.g., the type of faulty quantum gate). \qmutbench provides quantum software developers and testers with an accessible online dataset to obtain benchmarks of mutants necessary to assess either the quality of the test cases generated by their testing technique or to compare different testing techniques. It also enables the development of new mutation-guided quantum software testing techniques.\\
\noindent Online Interface: \url{https://enautmendi.github.io/QMutBench/}\\
Code \& Data: \url{https://github.com/EnautMendi/QMutBench}\\
Video: \url{https://youtu.be/6rjmiR16hdc} \\
\end{abstract}


\section{Introduction}
Quantum software promises to solve complex problems faster than classical software by utilising quantum phenomena such as superposition and entanglement. Such phenomena present new challenges for quantum software testing~\cite{qseRoadmapTOSEM2025}. Thus, several quantum software testing techniques have emerged~\cite{quantumTesting2030TOSEM, pauliStringsASE2024, testingQuantumICST2021}. \looseness=-1

As the field of quantum software testing grows, it is essential to have systematic methods to assess the effectiveness of tests generated by these techniques.
Typically, tests are assessed using benchmarks of faulty programs to measure their fault detection capability.
However, in this new field, benchmarks of real faulty quantum programs remain scarce~\cite{zhao2021bugs4q, qbugs}.
To address this limitation, mutation analysis, a technique that introduces artificial faults to programs to create their faulty versions, is often used to assess test effectiveness. 

In the context of quantum software testing, specialised tools have been developed for quantum-specific mutation analysis~\cite{Mendiluze2021,QmutPy}. 
Given a set of quantum programs (specified as {\it quantum circuits}), these tools can automatically generate a large number of {\it quantum circuit mutants} for each original quantum circuit.
However, these tools lack an automated and systematic method, supported by strong empirical evidence, to select a representative subset of mutants from all those generated.

In this data showcase, we address the challenge mentioned above, and we introduce \qmutbench, a dataset of quantum circuits accessible via a selection interface. We provide a large set of quantum circuit mutants and allow users to apply various selection criteria to obtain a benchmark of mutants that meet their specific requirements.

\qmutbench is built based on the result of an extensive empirical evaluation on quantum circuit mutants~\cite{MendiluzeUsandizaga2025}. The empirical evaluation was conducted on over 300 quantum circuits and over 700,000 quantum circuit mutants. The evaluation examined how these mutants interact with quantum circuit characteristics and their effects on \survivalrate, a metric quantifying the likelihood of survival (i.e., not being detected) for a mutant with specific characteristics. While that empirical evaluation offers valuable insights and a dataset of mutants for future research, its complexity limits the dataset's usability. \qmutbench addresses this issue by offering users filtering options to select a subset of mutants matching their criteria. We refer to the subset of mutants obtained from the full dataset as a {\it benchmark}, as it is the set of mutants used to assess test effectiveness. 

\qmutbench is implemented as an online interface that users can easily access to select a subset of mutants that meet their selection criteria (e.g., type of quantum algorithm, desired \survivalrate, or specific quantum gates used in the mutants). The selected subset of mutants together with the original quantum circuit are provided as a set of open quantum assembly language (\qasm) files~\cite{qasm}. This format is supported by most quantum computing platforms and tools, allowing for straightforward conversion to their respective programming languages.

Researchers could use the dataset to evaluate and improve new techniques for testing quantum software. Three main use cases that we identify are:
%
\begin{inparaenum}[(i)]
\item First, the benchmarks obtained from \qmutbench can be used to assess the quality of tests generated by a specific quantum software testing technique.
\item Second, the benchmarks acquired from \qmutbench can assist in developing new mutation-guided quantum testing techniques.
\item Third, the benchmarks provided by \qmutbench can be used to evaluate the effectiveness of quantum debugging techniques in locating and repairing faults.
\end{inparaenum}

\section{Background}\label{sec:Background}
This section introduces some concepts of quantum computing and quantum mutation analysis techniques.

\subsection{Quantum Computing}
Quantum computing uses distinct characteristics, such as {\it superposition} and {\it entanglement}, to perform computations that can potentially perform certain tasks faster than classical computing.
Quantum superposition is a key feature that enables computational speed-up as it allows a {\it quantum bit} ({\it qubit}) to exist in multiple states at the same time, in contrast to a classical bit, which can only exist in one state at a time (i.e., either 0 or 1)~\cite{yanofsky2008quantum}.
Quantum entanglement allows qubits to be correlated such that the state of one qubit is linked to the state of another qubit and a change in one will instantly affect the other~\cite{yanofsky2008quantum}. 

Quantum computers are programmed using {\it quantum circuits}, which consist of several qubits and {\it quantum gates} that perform operations on those qubits. Fig.~\ref{fig:quantumCircuit} shows a visual representation of a quantum circuit drawn in the IBM Quantum Composer~\cite{IBMQuantumComposer}.
\begin{figure}[!tb]
\centering
\includegraphics[width=1\linewidth]{QuantumCircuitWIthOperators.png}
\caption{An example of a quantum circuit alongside examples of the applications of mutation operators.
The circuit has four qubits ($q[0]$ to $q[3]$) and a group of four classical bits (denoted collectively as $c_{4}$). The circuit also shows examples of single-qubit gates (e.g., the Hadamard gate) and multi-qubit gates (e.g., the controlled-NOT gate). Mutations introduced by quantum mutation operators are highlighted in orange, along with the positions of the gates in the circuit. For instance, the \textit{Remove} mutation operator removes the controlled-NOT gate to create a mutant of the original circuit.}
\label{fig:quantumCircuit}
\end{figure}
%
The circuit consists of four qubits, $q[0]$ to $q[3]$, and four classical bits, collectively represented as $c_4$. A sequence of quantum gates is applied to the qubits. For example, a single-qubit Hadamard gate (denoted as H) is applied to $q[0]$, placing it into an equal superposition of the states 0 and 1. A multi-qubit gate (the controlled-NOT (CNOT)) is applied between $q[0]$ and $q[1]$, where the value of $q[0]$ determines whether the state of $q[1]$ is to be flipped. The measurement operations in Fig.~\ref{fig:quantumCircuit} measure the state of the qubits and store the measured values into classical bits. These operations collapse the quantum superpositioned states into definite classical states. 

\subsection{Quantum Mutation Analysis}\label{subsec:backPrevStudy}
Mutation analysis is widely used in classical software testing~\cite{MutationSurvey2} to assess the effectiveness of tests in detecting faults. To achieve this, it employs various {\it mutation operators} to introduce small changes ({\it mutations}) in a program and create faulty versions of the program under test ({\it mutants}). The tests are then executed on these mutants to determine if they can detect them. The effectiveness of the tests is typically evaluated based on the proportion of mutants they detect, which is measured using a metric called the {\it mutation score}.

In quantum mutation analysis, the programs under test are quantum circuits, which require the definition of specific mutation operators. Some quantum mutation tools have already been developed for this purpose~\cite{QmutPy,Mendiluze2021}. These tools rely on gate-based operations to generate mutants, typically by adding, removing, or modifying the gates in a quantum circuit. Fig.~\ref{fig:quantumCircuit} shows an example of a quantum circuit and examples of three possible mutations in the circuit (shown in orange) by applying three mutation operators, i.e., \textit{Add}, \textit{Remove}, and \textit{Replace}. The {\it Add} operator introduces a new gate to the circuit, the {\it Remove} operator removes an existing gate from the circuit, and the {\it Replace} operator substitutes an existing gate with another gate. For example, in the figure, a two-qubit gate---specifically, a controlled-NOT between $q[1]$ and $q[2]$---is removed to create a mutant of the original circuit.


\section{\qmutbench}\label{sec:tool}

\qmutbench features an online dataset containing 375 original quantum circuits and their corresponding mutants, totalling 723,079. 
\qmutbench enhances the accessibility of mutants to the public, enabling researchers and developers to systematically access and retrieve only the mutants relevant to their needs, without downloading the entire dataset.

\subsection{Data Source and Methodology}
The mutants that comprise the \qmutbench dataset were obtained by executing a previously developed quantum mutation tool called Muskit~\cite{Mendiluze2021}. We used Muskit to generate all mutants from our selected original quantum circuits; we generated all mutant combinations by applying all three operators in all supported quantum gates, and across all positions in the circuit. In a previous study~\cite{MendiluzeUsandizaga2025}, we explored how the characteristics of quantum circuits and algorithms affect mutant detection and defined the \survivalrate as a metric for assessing the likelihood of a mutant's survival. This rate is calculated for each group of mutants having the same characteristics (i.e., algorithm, number of qubits, mutation operator, gate affected by the mutation, and position interval of the mutation in the circuit). It is computed in an opposite way to the mutation score: instead of considering the percentage of mutants that are detected, it considers the percentage of mutants that are not detected. We chose this metric as it puts more emphasis on the difficulty to detect a mutant, which is important when building benchmarks of faulty quantum circuits.

Our previous study~\cite{MendiluzeUsandizaga2025} resulted in an extensive dataset of quantum circuit mutants that is hard to retrieve and understand, making it impractical due to the large number of mutants. The distribution of the data's characteristics is reported in~\cite{MendiluzeUsandizaga2025}. 

\begin{figure}[!tb]
\centering
\includegraphics[width=0.9\linewidth]{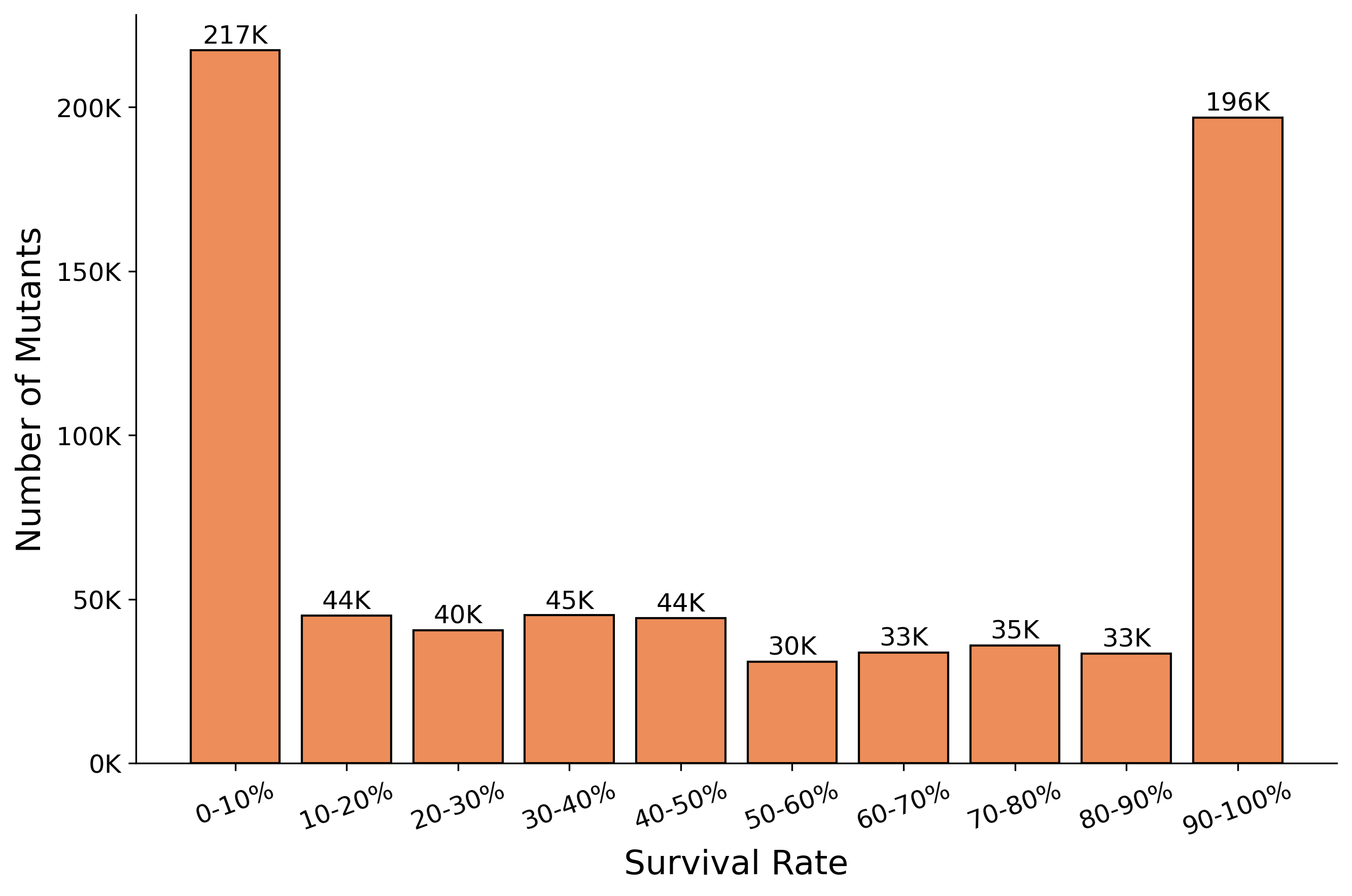}
\caption{Distribution of mutants across \survivalrate ranges}
\label{fig:SRDist}
\end{figure}

In this work, we have adapted the data to make it more accessible for researchers and practitioners. We grouped all the mutants by their characteristics and labelled them by the \survivalrate of the group they belong to, enabling us to retrieve the required mutants as needed. Fig.~\ref{fig:SRDist} shows new descriptive data representing the number of mutants that belong to groups with a given \survivalrate.
One can directly see from the figure that the dataset mostly contains mutants in groups with \survivalrate of 0-10\% or 90-100\% (i.e., mutants that are either very easy to detect or very hard to detect and equivalent), while still showing a representative number of mutants across other \survivalrate values.

We developed an online interface that enables users to filter the dataset and select mutants relevant to their needs. Users can then download only the selected mutants without needing to manually retrieve them from the extensive dataset.
Furthermore, we translated all the quantum circuits and mutants into \qasm, which can be considered the de facto standard language for quantum circuits. In the previous study, all quantum circuits, both original and mutated, were written only for Qiskit; this change now allows users to execute them in their preferred quantum programming language.

\subsection{Usage}\label{sec:usage}
\begin{figure}[!tb]
\centering
\includegraphics[width=0.9\linewidth]{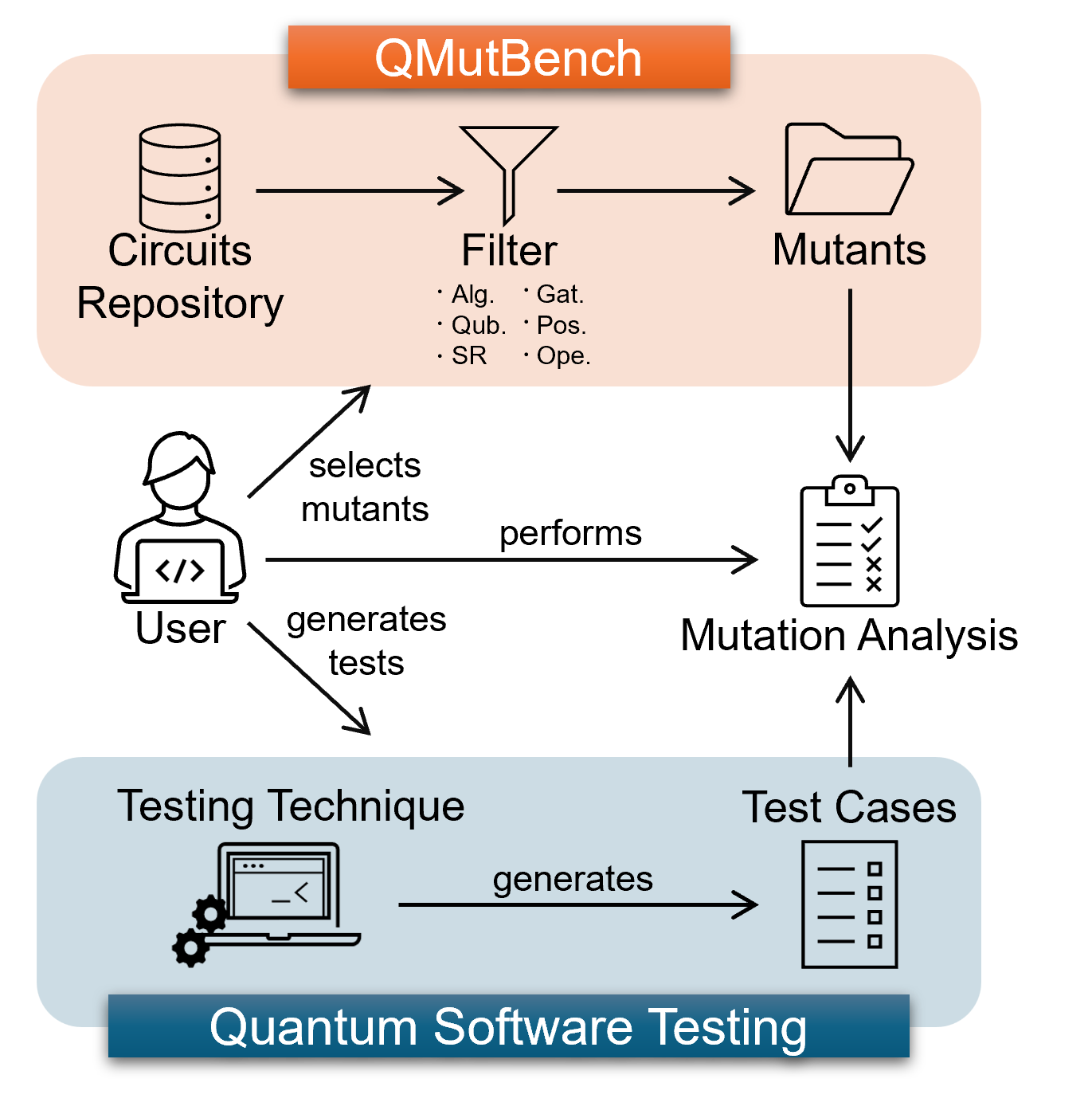}
\caption{Flow of usage of \qmutbench}
\label{fig:userflow}
\end{figure}

Fig.~\ref{fig:userflow} illustrates the process that a user will follow in the case where \qmutbench is used to assess the quality of tests generated by a specific quantum software testing technique. 
On one hand, the user will have access to the tests they intend to assess. 
The assessment can have different purposes (only one possibility is shown in Fig.~\ref{fig:userflow}), such as:
\begin{inparaenum}[(i)]
\item evaluating the quality of the tests obtained with a given testing technique;
\item comparing the effectiveness of different testing techniques; or
\item assessing a given set of tests (not generated with any particular technique).
\end{inparaenum}

On the other hand, the user will have access to our online interface that provides access to the quantum circuit mutants dataset.
By applying filtering operations, the user can obtain a benchmark of quantum circuit mutants along with their corresponding original circuits.

This filtering allows the user to select specific mutants based on various criteria, such as \survivalrate, to evaluate the effectiveness of tests in detecting mutants with varying levels of difficulty. Moreover, by selecting specific mutation characteristics, the user can focus on aspects, such as positions within the circuit, or identify tests specialised in detecting faults associated with specific quantum gates.


\subsection{\qmutbench online interface}\label{sec:onlineInterface}
\begin{figure}[!tb]
\centering
\includegraphics[width=0.95\linewidth]{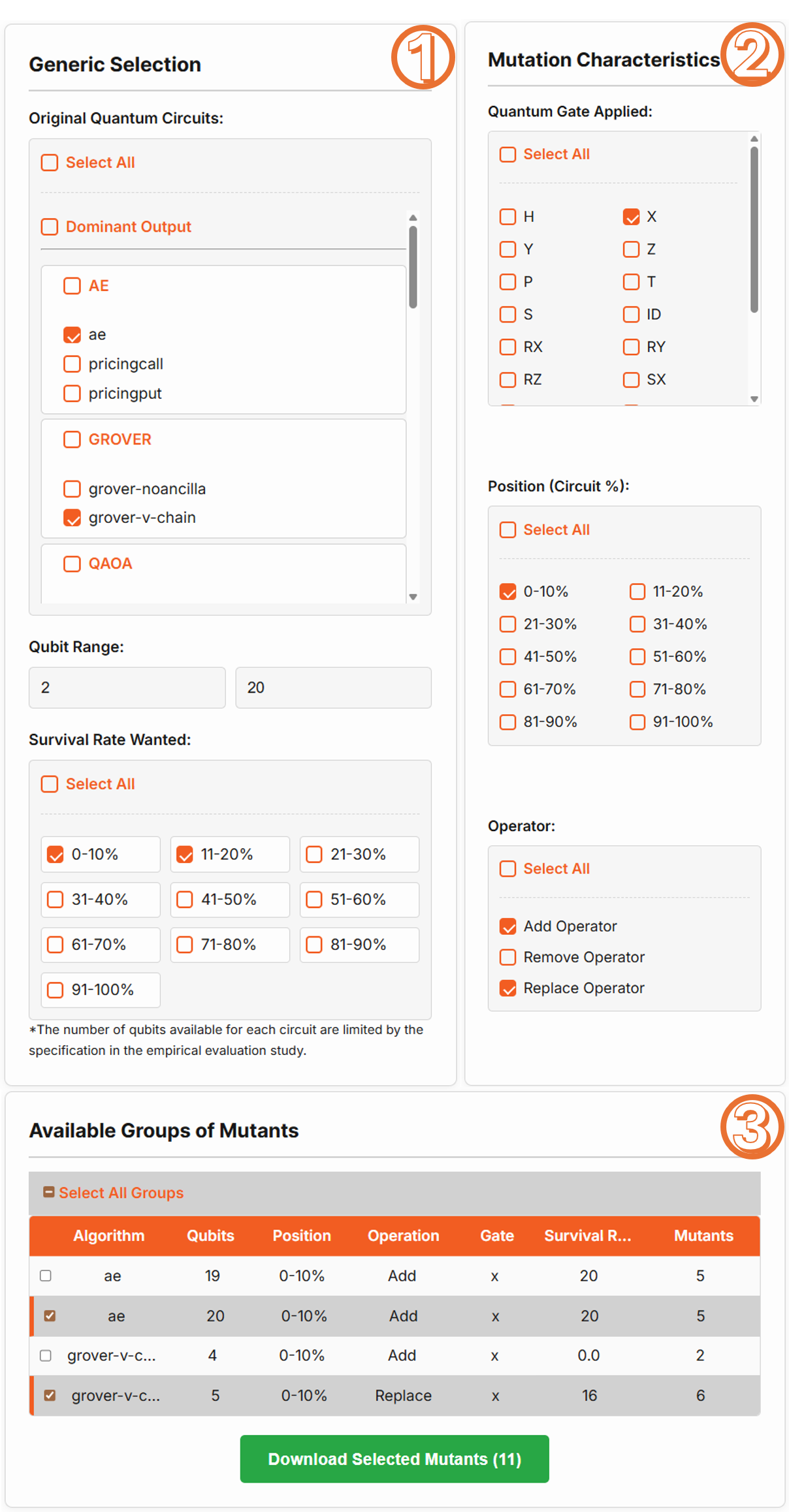}
\caption{\qmutbench online interface}
\label{fig:webpage}
\end{figure}

Fig.~\ref{fig:webpage} shows the three panels of the online interface that guide the user through the dataset, enabling an easier retrieval of mutants.

\subsubsection{Panel ``Generic Selection''}
The first panel of \qmutbench shown in Fig.~\ref{fig:webpage} is the only one that is enabled at the start. This panel contains the generic selection fields that users need to select before obtaining the desired benchmark of mutants.

In the first checkbox group ({\tt Original Quantum Circuits}), all the quantum algorithms are listed, divided into two columns based on the output type: {\it dominant} or {\it non-dominant}. Dominant output algorithms aim to identify the quantum state with the highest probability, while non-dominant output algorithms focus on the output distribution of several output states. The original quantum algorithms are then further categorised into subgroups based on the base algorithm (Amplitude Estimation (AE), Grover, etc.). This is a fundamental step for selecting the desired algorithms that will constitute the benchmark of quantum circuits the user downloads from the dataset.

Next, the user can choose the desired number of qubits ({\tt Qubits Range}) of the quantum circuits to retrieve. \qmutbench contains circuits having from 2 to 30 qubits; however, not all algorithms support every possible qubit count within that range.

The checkbox group {\tt Survival Rate Wanted} allows the user to choose different \survivalrate ranges for the mutants to select.

Once the desired original circuits, qubit range, and the specific \survivalrate for the mutants are selected, the remaining two panels of the tool will be enabled.

\subsubsection{Panel ``Mutation Characteristics''}
This panel of the interface allows users to specify the types of mutations to include by selecting specific mutation characteristics. 
If no characteristic is selected, all mutants will be selected by default. 
There are three main mutation categories: {\tt Quantum Gate Applied}, {\tt Position}, and {\tt Operator}.

The {\tt Quantum Gate Applied} panel lists all the quantum gates that were applied to create mutants from the original circuits. A user can choose one or more of the listed quantum gates; alternatively, the user can apply {\tt Select all} to select all the gates. For example, if \textit{H} gate is selected, the mutants where this gate is added, removed, or replaced are shown.

A {\tt Position} in a quantum circuit refers to the location where mutations were introduced. These positions are categorised as percentages of the total possible positions in the circuit. For example, selecting the 0-10\% range will return only those mutants where mutations were introduced within the first 10\% of the circuit's positions. 


An \textit{Operator} determines which mutation operator was applied to create a mutant. There are three operators, i.e., \textit{Add}, \textit{Remove}, and \textit{Replace} gate. For example, if the \textit{Add} operator is selected, the interface will only show mutants where the \textit{Add} operator was applied to create the mutants. 

Naturally, some combinations of the characteristics that meet the user's needs may not result in mutants with the desired \survivalrate. In such cases, the interface will display no available groups of mutants for download. 

\subsubsection{Panel ``Available Groups''}\label{subsec:dowloadMutants}
In the last panel of the interface ({\tt Available Groups}), the resulting groups of mutants are displayed. These groups are updated whenever a new original circuit, \survivalrate range, or characteristic is selected. They represent groups of mutants with different characteristics that match the user's filters. Each group may contain several mutants, which represent different positions within the specified range.
If at least one group of mutants is selected, the download button will display the total number of quantum circuit mutants to be downloaded.

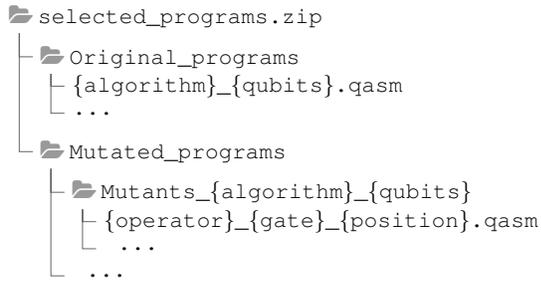
\begin{figure}[!tb]
\centering


\newcommand{\FTdir}{}
\def\FTdir(#1,#2,#3){%
  \FTfile(#1,{{\color{black!40!white}\faFolderOpen}\hspace{0.2em}#3})
  (tmp.west)++(0.8em,-0.4em)node(#2){}
  (tmp.west)++(1.5em,0)
  ++(0,-1.3em) 
}

\newcommand{\FTfile}{}
\def\FTfile(#1,#2){%
  node(tmp){}
  (#1|-tmp)++(0.6em,0)
  node(tmp)[anchor=west,black]{\tt #2}
  (#1)|-(tmp.west)
  ++(0,-1.2em) 
}
\newcommand{\FTroot}{}
\def\FTroot{tmp.west}
\centering
\resizebox{0.85\columnwidth}{!}{
\begin{tikzpicture}%
\draw[color=black!60!white]
  \FTdir(\FTroot,root,selected\_programs.zip){
      ++(0,-0.5em) 
    \FTdir(root,Originprograms,Original\_programs){       
      \FTfile(Originprograms,\{algorithm\}\_\{qubits\}.qasm)
      \FTfile(Originprograms,\ldots)
    }
    ++(0,-0.5em) 
    \FTdir(root,Mutatedprograms,Mutated\_programs) {
        ++(0,-0.5em) 
      \FTdir(Mutatedprograms,MutantsalgorithmXqubits,Mutants\_\{algorithm\}\_\{qubits\}){
        \FTfile(MutantsalgorithmXqubits,\{operator\}\_\{gate\}\_\{position\}.qasm)
        \FTfile(MutantsalgorithmXqubits, \ldots)
      }
      \FTfile(Mutatedprograms, \ldots)}
  };
\end{tikzpicture}}
\caption{Structure of the downloaded folder}
\label{fig:foldertree}
\end{figure}

After pressing the download button, a {\tt .zip} file will be downloaded, which will include the selected original quantum circuits and their respective mutants as \qasm files ({\tt .qasm}). This {\tt .zip} file is structured as shown in Fig.~\ref{fig:foldertree}.
The downloaded file contains two main folders: one for the original circuits and another for the mutated circuits. The ``mutants'' folder is further divided into subfolders organised by the selected algorithm and qubit settings. Each file is named to indicate the operator used, the modified gate, and the position of the change in the filename, as done by some classic mutation analysis tool~\cite {ma2005mujava}. The original quantum circuits and folders are named based on the algorithms they use and the number of qubits they contain.

\section{Extendability and Future Work}\label{sec:extension}
\qmutbench can be easily extended by adding new original quantum circuits and their corresponding mutants to the GitHub repository, following the same naming conventions described above in Sect.~\ref{subsec:dowloadMutants}. Relevant data, including program and mutant characteristics and \survivalrate, should also be updated in the provided .csv files. The interface need not be modified unless a new characteristic or an entirely new type of program is introduced. As future work, we plan to integrate this process as part of the online interface, where a user could add the original quantum circuit and the respective mutants directly to the repository, and the .csv files would update automatically. To further improve the dataset, we also aim to explore integrating AI-driven recommendation systems, such as large language models. If the user has a quantum circuit that is not present in the dataset, the recommendation system could identify similar patterns in known quantum circuits and suggest mutations to apply, based on the survival rates observed in those circuits. 

\qmutbench enables several directions for future research. 
In the context of mutation analysis, researchers could investigate research questions where the mutants are analysed under a noisy environment of real quantum devices, such as: ``How do mutant \survivalrate change when executed on noisy simulators and on real quantum hardware?'' or ``Which mutant characteristics are most relevant for detection in noisy devices?''. 
Researchers could also simply use the dataset to assess the new quantum software testing tools they develop or to compare different approaches, as mentioned earlier in Sect.~\ref{sec:usage}. 

Furthermore, quantum software debugging and repair could also benefit highly from \qmutbench as the mutants provided help in evaluating the efficiency of these techniques in detecting and repairing quantum faults. 


\section{Conclusion}\label{sec:conclusion}
We introduced \qmutbench, a dataset with a selection interface for quantum circuit mutants. The dataset includes over 300 original quantum circuits and over 700,000 mutants, all of which are available online. It provides several selection criteria to help users easily select and download mutants, starting from the original quantum circuit(s) and focusing on specific mutation characteristics. Furthermore, users can choose mutants based on their \survivalrate. The selected mutants can then be used to assess the effectiveness of the tests. By making this dataset publicly available, we ensure that the largest collection of quantum circuit mutants is accessible to the broader quantum software community.

\bibliographystyle{IEEEtran}
\bibliography{References}
\end{document}